\documentclass[aps,prd,twocolumn,groupedaddress,showpacs]{revtex4}
\usepackage{float,graphicx}
\begin{document}

% Use the \preprint command to place your local institutional report
% number in the upper righthand corner of the title page in preprint mode.
% Multiple \preprint commands are allowed.
% Use the 'preprintnumbers' class option to override journal defaults
% to display numbers if necessary
%\preprint{}

%Title of paper
\title{Big Bang Nucleosynthesis Constraints on Universal
Extra Dimensions and Varying Fundamental Constants}

% repeat the \author .. \affiliation  etc. as needed
% \email, \thanks, \homepage, \altaffiliation all apply to the current
% author. Explanatory text should go in the []'s, actual e-mail
% address or url should go in the {}'s for \email and \homepage.
% Please use the appropriate macro foreach each type of information

% \affiliation command applies to all authors since the last
% \affiliation command. The \affiliation command should follow the
% other information
% \affiliation can be followed by \email, \homepage, \thanks as well.

\author{B.~Li }
\email[Email address: ]{bli@phy.cuhk.edu.hk}
%\homepage[]
%\thanks{}
%\altaffiliation{}
\affiliation{Department of Physics, The Chinese University of Hong
Kong, Hong Kong SAR, China}

\author{M.~-C.~Chu}
\email[Email address: ]{mcchu@phy.cuhk.edu.hk}
%\homepage[]
%\thanks{}
%\altaffiliation{}
\affiliation{Department of Physics, The Chinese University of Hong
Kong, Hong Kong SAR, China}

%Collaboration name if desired (requires use of superscriptaddress
%option in \documentclass). \noaffiliation is required (may also be
%used with the \author command).
%\collaboration can be followed by \email, \homepage, \thanks as well.
%\collaboration{}
%\noaffiliation

\date{\today}

\begin{abstract}
The successful prediction of light element abundances from Big
Bang Nucleosynthesis (BBN) has been a pillar of the standard model
of Cosmology. Because many of the relevant reaction rates are
sensitive to the values of fundamental constants, such as the fine
structure constant and the strong coupling constant, BBN is a
useful tool to probe and to put constraints on possible
cosmological variations of these constants, which arise naturally
from many versions of extra-dimensional theories. In this paper,
we study the dependences of fundamental constants on the radion
field of the universal extra dimension model, and calculate the
effects of such varying constants on BBN. We also discussed the
possibility that the discrepancy between BBN and the Wilkinson
Microwave Anisotropy Probe (WMAP) data on the baryon-to-photon
ratio can be reduced if the volume of the extra dimensions was
slightly larger - by $\mathcal{O} (10^{-3})$ - at the BBN era
compared to its present value, which would result in smaller gauge
couplings at BBN by the same factor.
\end{abstract}

% insert suggested PACS numbers in braces on next line
% insert suggested keywords - APS authors don't need to do this
%\keywords{}

%\maketitle must follow title, authors, abstract, \pacs, and \keywords
\maketitle

% body of paper here - Use proper section commands
% References should be done using the \cite, \ref, and \label commands

\section{INTRODUCTION}

The standard Big Bang Nucleosynthesis (BBN) theory is a successful
marriage between standard Friedmann cosmology and nuclear physics,
explaining the origin and abundances of the light elements D,
$^{4}$He, $^{3}$He and $^{7}$Li. Its only one input parameter, the
baryon-to-photon ratio $\eta = n_{b}/n_{\gamma}$ (where $n_{b}$
and $n_{\gamma}$ are respectively the number density of baryons
and photons) has now been determined by the observations of the
Wilkinson Microwave Anisotropy Probe (WMAP) with rather good
accuracy \cite{Spergel2003}, so that there are essentially no free
parameters in this scenario. Because the BBN predictions are
sensitive to a set of physical quantities which depend on various
fundamental constants such as the gauge couplings, the Yukawa
couplings and gravitational constant (see Sections III and IV for
details), it would provide stringent constraints on the
cosmological variations of these constants. For example, in
\cite{BIR,Avelino2001,NL} the authors use BBN to constrain the
change of fine structure constant, in \cite{Dixit1988,
Scherrer1993, Yoo2003} constraints on the change of Higgs vacuum
expectation value are considered, and the effects of a varying
strong coupling are discussed in \emph{e.g.} \cite{Flambaum2002,
Kneller2003}; the variation of gravitational constant has also
been investigated \cite{Casas1992ab, Serna1992, Santiago1997,
Damour1999, Clifton2005} in the context of BBN. Besides, there is
another reason for the interests in the interplay between varying
fundamental constants and BBN: in contrast to the excellent
consistency between BBN theory and observation with $\eta$ given
by WMAP for the deuteron (D) abundance, the predicted $^{4}$He and
$^{7}$Li abundances are smaller than the results implied by WMAP
\cite{Cyburt2003, Steigman2003}; it is then suggested that the
variations of fundamental constants such as the fine structure
constant \cite{NL, Ichikawa2004} or the deuteron binding energy
$B_{d}$ \cite{Dmitriev2004} ($B_{d}$ itself is certainly not a
fundamental constant, but its variation possibly originates from
some other fundamental constants as suggested in \cite{Yoo2003})
\emph{etc}. might partly or completely solve this discrepancy.

The theoretical investigations of varying fundamental constants
date back to the early work of Dirac in 1930's, and have
re-aroused great attentions because of the recent discovery of
Webb \emph{et al.} \cite{Webb1999, Murphy2001} that the quasar
absorption lines at redshifts of $z$ = 1 $\sim$ 3 suggest a small
evolution of the fine structure constant between that period and
present. Although variations of the fundamental constants do not
occur in the standard model (SM) of particle physics, the two
leading paradigms of the physics beyond SM, namely the string
theory \cite{Polchinski1998} and the Kaluza Klein theory
\cite{Overduin1997}, generally predict such variations. In these
scenarios a dilaton field or the size of extra dimensions may
evolve cosmologically; thus until their vacuum expectation values
(VEV) are fixed, there might be a co-variation of several
fundamental constants. This picture is similar to the discussions
in grand unified theories (GUTs) \cite{Langacker2002, Calmet2002}
and implies that considerations of one varying constant alone may
be incomplete, although indeed such a co-variation itself is in
general rather model-dependent (see \cite{Campbell1995,
Ichikawa2002} for one case derived from string theory). For a
detailed discussion of the theoretical and observational
(experimental) aspects of varying fundamental constants, see
\cite{Uzan2003}.

The recent interests in the extra dimensional theories have
stimulated other considerations of varying fundamental constants.
Of the frequently discussed extra dimension models, the original
ADD \cite{ADD1999} and RS \cite{RS1999} brane models do not induce
changes in the gauge couplings even though their moduli fields
evolve \cite{Brax2003a, Brax2003b} because of the conformal
invariance of the gauge kinetic terms (see however
\cite{Palma2003} for an alternative), and so in this work we will
concentrate on the universal extra dimension (UED) model, in which
the fields can propagate in all dimensions. This model was first
proposed in \cite{Appelquist2001} and became extremely interesting
for cosmology since it was found later that it predicted the
presence of a stable massive particle, the lightest KK partner
(LKP), which is a natural dark matter candidate (see
\cite{Servant2002, Kong2005} and references therein for details).
Our purpose in this work is twofold: firstly, we study how the
fundamental constants in the low energy effective theory change if
the size of the universal extra dimensions, the radion field,
undergoes a slow cosmological evolution and to what extent the
changes are allowed by the BBN observations; secondly, we show
that if the volume of the extra space at the time of BBN is
slightly larger than its present value, the discrepancy between
BBN and WMAP discussed above may be reduced.

The arrangement of this work is the following: in Section II we
derive the effective low energy actions for the gravitational and
matter sectors in the model; then we shall find the radion
dependences of the fundamental constants in Section III. In
Section IV we briefly review the standard BBN, point out the most
relevant physical quantities for its prediction and relate these
quantities to the radion field.  Section V contains the numerical
results of this work, which are obtained by modifying the standard
code of BBN \cite{Kawano1992} and including the effects discussed
in Section IV. Finally Section VI is devoted to discussion and
conclusion. Throughout this work we assume 3 species of massless
neutrinos as in the standard BBN and adopt the units $\hbar=c=1$.

\section{THE LOW ENERGY 4-DIMENSIONAL EFFECTIVE ACTIONS}

Let us consider a general $4+n$ dimensional model, with $n$ being
the number of the (universal) extra dimensions (although there may
also be large extra dimensions, we do not consider them in this
work). The full line element is given as:
\begin{equation}
ds^{2} =
G_{AB}dX^{A}dX^{B}=g_{\mu\nu}dx^{\mu}dx^{\nu}+h_{ab}dy^{a}dy^{b},
\end{equation}
where $\mu$, $\nu$ = 0, 1, 2, 3 label the four ordinary
dimensions, $a$, $b$ = $4, \cdot\cdot\cdot, 3+n$ denote extra
dimensions and $A, B = 0, 1, 2, \cdot\cdot\cdot, 3+n$ the whole
spacetime. For simplicity we shall not consider cross terms such
as $G_{a\mu}$ in Eq.~(1). The extra dimensions are assumed to
compactify on an orbifold, and their coordinates $y_{a}$ take
values in the range [0, 1]. The quantities $h_{ab}$ have
dimensions of [Length]$^{2}$ since $y_{a}$ are dimensionless in
our choice.

Because the energy range we are interested in is much lower than
the inverse size of the universal extra dimensions, which is
thought to be larger than several hundred GeV's, it is adequate to
consider only the zero modes of the metric. Then the effective
4-dimensional action (in the gravitational sector) can be obtained
by dimensionally reducing Eq.~(1) as:
\begin{widetext}
\begin{eqnarray}
S_{\mathrm{Gravity}} &=& \frac{1}{\kappa_{4+n}^{2}}\int
d^{4+n}X\sqrt{|G|}R_{4+n}[G]\nonumber\\ &=&
\frac{1}{\kappa_{4}^{2}}\int d^{4}xd^{n}y \sqrt{|g|}
\frac{\sqrt{|h|}}{V_{0}}\left[R_{4}\left[g\right]-
\frac{1}{4}\partial_{\mu}h^{ab}\partial^{\mu}h_{ab}
-\frac{1}{4}h^{ab}\partial_{\mu}h_{ab}\cdot
h^{cd}\partial^{\mu}h_{cd}\right],
\end{eqnarray}
\end{widetext}
in which $|g|$, $|h|$ and $|G|$ are respectively the determinants
of the metrics of the ordinary dimensions, the extra dimensions
and the whole spacetime. $R_{4}[g]$ and $R_{4+n}[G]$ are the Ricci
scalars of the ordinary 4 and the total $4+n$ dimensional
spacetimes. $\kappa_{4}, \kappa_{4+n}$ are related to the 4 and
$4+n$ dimensional Planck masses via $\kappa_{4}^{2} = 2 M_{4}^{2}$
and $\kappa_{4+n}^{2} = 2 M_{4+n}^{2+n}$, while they themselves
are connected by a volume suppression $\kappa_{4+n}^{2} =
\kappa_{4}^{2}\cdot V$, with $V$ being a measure of the extra
space volume whose present-day value is denoted by $V_{0}$ in
Eq.~(2) (Note that because of the specified choice of $V_{0}$ and
because the higher dimensional quantity $\kappa_{4+n}$ is treated
as a constant, the $\kappa_{4}$ above also takes its currently
measured value and is a constant rather than a variable).

The effective 4-dimensional curvature term is not canonical in
Eq.~(2); to make it so, let us take the conformal transformation
\begin{equation}
g_{\mu\nu} \rightarrow e^{2\phi} g_{\mu\nu}
\end{equation}
and choose the field $\phi$ to satisfy
\begin{equation}
\frac{\sqrt{|h|}}{V_{0}}e^{2\phi} = 1.
\end{equation}
Then we obtain the effective 4-dimensional gravitational action in
the Einstein frame:
\begin{widetext}
\begin{equation}
S_{\mathrm{Gravity}} = \frac{1}{\kappa_{4}^{2}}\int
d^{4}x\sqrt{|g|}
\left[R_{4}-\frac{1}{4}\partial_{\mu}h^{ab}\partial^{\mu}h_{ab}
+\frac{1}{8}h^{ab}\partial_{\mu}h_{ab}\cdot
h^{cd}\partial^{\mu}h_{cd}\right].
\end{equation}
\end{widetext}

We shall make a further assumption that the extra dimension(s) are
homogeneous and isotropic, \emph{i.e}. the metric of the extra
space takes the following form:
\begin{equation}
h_{ab} = \mathrm{diag} (-b^{2}, -b^{2}, \cdot\cdot\cdot, -b^{2}),
\end{equation}
and then the action Eq.~(5) could be rewritten as
\begin{equation}
S_{\mathrm{Gravity}} = \int
d^{4}x\sqrt{|g|}\left[\frac{1}{\kappa_{4}^{2}}R_{4}+\frac{1}{2}
g^{\mu\nu}\partial_{\mu}\sigma\partial_{\nu}\sigma\right]
\end{equation}
by defining a new scalar field, the radion $\sigma$:
\begin{equation}
\sigma \equiv
\frac{1}{\kappa_{4}}\sqrt{\frac{n+2}{n}}\mathrm{log}\frac{b^{n}}{V_{0}}.
\end{equation}

Now we turn to the matter sector of the effective 4-dimensional
action, first considering the scalar fields. The action of a $4+n$
dimensional scalar field $\tilde{\varphi}$ (all the quantities
with tilde are higher dimensional in this work) is given by:
\begin{equation}
S_{\tilde{\varphi}} = \int d^{4+n}X\sqrt{|G|}
\left[\frac{1}{2}G^{AB}\partial_{A}\tilde{\varphi}
\partial_{B}\tilde{\varphi}-\tilde{U}(\tilde{\varphi})\right].
\end{equation}
Since we are only considering the zero-mode theory and the
zero-modes of the fields are independent of the extra dimensional
coordinates (see Appendix A), we define a 4-dimensional scalar
field (which has the correct dimension) $\varphi$ from
$\tilde{\varphi}$:
\begin{equation}
\varphi (x) \equiv \sqrt{V_{0}}\tilde{\varphi} (x, y).
\end{equation}
With this new field, the action Eq.~(9) could be rewritten as
\begin{equation}
S_{\varphi} = \int d^{4}x\sqrt{|g|}\frac{\sqrt{|h|}}{V_{0}}
\left[\frac{1}{2}g^{\mu\nu}\partial_{\mu}\varphi
\partial_{\nu}\varphi-U(\varphi)\right],
\end{equation}
where the new potential is related to the old one by
\begin{equation}
U (\varphi) \equiv V_{0}\tilde{U}(\tilde{\varphi}).
\end{equation}
Then the same conformal transformation Eqs.~(3) and (4) transforms
Eq.~(11) into the following canonical form of the effective
action:
\begin{widetext}
\begin{equation}
S_{\varphi} = \int d^{4}x\sqrt{|g|}
\left\{\frac{1}{2}g^{\mu\nu}\partial_{\mu}\varphi
\partial_{\nu}\varphi-\mathrm{exp}\left[-\kappa
\sqrt{\frac{n}{n+2}}\sigma\right]U(\varphi)\right\}.
\end{equation}
\end{widetext}
Note that from now on we will use $\kappa$ instead of $\kappa_{4}$
for simplicity.

The same technique could be applied to gauge fields, whose higher
dimensional action is given as
\begin{equation}
S_{\mathrm{Gauge}} = -\int d^{4}xd^{n}y\sqrt{|g|}\sqrt{|h|}
\frac{1}{4\tilde{g}^{2}}\tilde{F}^{rAB}\tilde{F}_{AB}^{r},
\end{equation}
where $\tilde{g}$ is the $4+n$ dimensional gauge coupling constant
and $\tilde{F}_{AB}^{r}$ are the corresponding gauge field
strengths. Taking the following redefinitions of the zero-mode
gauge field
\begin{equation}
F_{\mu\nu}^{r} \equiv \sqrt{V_{0}}\tilde{F}_{\mu\nu}^{r}, \
F_{ab}^{r} \equiv 0,
\end{equation}
and the conformal transformation Eqs.~(3), (4), we finally obtain
the effective 4-dimensional action in the Einstein frame as:
\begin{equation}
S_{\mathrm{Gauge}} = -\int d^{4}x\sqrt{|g|}
\frac{1}{4\tilde{g}^{2}}\mathrm{exp}\left[\kappa
\sqrt{\frac{n}{n+2}}\sigma\right]F^{r\mu\nu}F_{\mu\nu}^{r}.
\end{equation}

One can obtain the effective action for the Dirac fermion field
similarly:
\begin{widetext}
\begin{equation}
S_{\Psi} = \int d^{4}x \sqrt{|g|}
\left\{\mathrm{exp}\left[-\frac{\kappa}{2}\sqrt{\frac{n}{n+2}}
\sigma\right]e_{d}^{\mu}\bar{\Psi}i\Gamma^{d}D_{\mu}\Psi
+\mathrm{exp}\left[-\kappa\sqrt{\frac{n}{n+2}}
\sigma\right]m\bar{\Psi}\Psi\right\},
\end{equation}
\end{widetext}
where $e_{d}^{\mu}$ is the vierbein and $\Gamma^{d}$ are the Dirac
matrices in the tangent space embedded in the $4+n$ dimensional
flat spacetime. The fermion mass $m = y\langle H\rangle$ is
acquired from the Higgs mechanism with the Yukawa coupling
constant $y$ and Higgs scalar field VEV $\langle H\rangle = v$,
which connect to their higher dimensional counterparts by $y =
\tilde{y}/\sqrt{V_{0}}$ and $v = \sqrt{V_{0}}\tilde{v}$.

To make the kinetic part of the fermion action Eq.~(17) canonical,
we rescale the field as
\begin{equation}
\psi \equiv
\mathrm{exp}\left[-\frac{\kappa}{4}\sqrt{\frac{n}{n+2}}\sigma\right]\Psi,
\end{equation}
and then by the conformality of the coupling of massless Weyl
fermions, Eq.~(17) becomes:
\begin{widetext}
\begin{equation}
S_{\psi} = \int d^{4}x \sqrt{|g|}
\left\{e_{d}^{\mu}\bar{\psi}i\Gamma^{d}D_{\mu}\psi
+\mathrm{exp}\left[-\frac{\kappa}{2}\sqrt{\frac{n}{n+2}}
\sigma\right]m\bar{\psi}\psi\right\}.
\end{equation}
\end{widetext}

Our results above are equal to those of \cite{Mazumdar2004} when
there are no large extra dimensions in their model.

\section{RADION DEPENDENCE OF FUNDAMENTAL CONSTANTS}

If there is a slow cosmological evolution of the extra dimensional
size between the time of BBN and now, then some or all of the
fundamental constants in the particle physics standard model will
be changed and these changes may alter the results of the standard
BBN. Because the standard BBN is sensitively dependent on
fundamental constants, these changes, if exist, would be
constrained stringently by BBN. Furthermore, they have the
potential of slightly modifying some aspects of the standard BBN
and improving the agreements between theoretical calculation and
observations. This section is devoted to how the fundamental
constants depend on the size of the extra dimensions (or equally
the radion field), and in the next section we shall consider how
BBN is influenced by these varying constants.

As we are considering the system in the Einstein frame, the
(4-dimensional) Planck mass will stay constant. Because only
dimensionless quantities such as the ratios of masses are
physically significant, we shall take the Planck mass as a
reference scale while expressing the variations of other
quantities with dimension of mass such as
$\Lambda_{\mathrm{QCD}}$.

The Higgs boson is a scalar field; therefore its radion
dependence, as described by Eq.~(13), is solely through an overall
rescaling of the radion potential. As a result, the Higgs VEV,
which is determined by minimizing the potential, will not be
modified by the evolution of the radion. We shall take it to be a
constant in the following calculation. Consequently, the radion
dependence of fermion masses, as indicated by Eq.~(19), should be
due to the radion dependence of the 4-dimenioanl Yukawa coupling
constant $y$. Let $y_{\mathrm{BBN}}$ and $y_{\mathrm{NOW}}$ denote
the Yukawa couplings at the time of BBN and now, and then from
Eq.~(19) we have:
\begin{widetext}
\begin{equation}
\frac{y_{\mathrm{BBN}}}{y_{\mathrm{NOW}}} =
\frac{\mathrm{exp}\left[-\frac{\kappa}{2}\sqrt{\frac{n}{n+2}}
\sigma_{\mathrm{BBN}}\right]}{\mathrm{exp}\left[-\frac{\kappa}
{2}\sqrt{\frac{n}{n+2}}\sigma_{\mathrm{NOW}}\right]} =
\mathrm{exp}\left[-\frac{\kappa}{2}\sqrt{\frac{n}{n+2}}
\sigma_{\mathrm{BBN}}\right] \equiv \rho^{-\frac{1}{2}},
\end{equation}
\end{widetext}
where we have used the definition $\rho \equiv
V_{\mathrm{BBN}}/V_{\mathrm{NOW}} = V_{\mathrm{BBN}}/V_{0}$. In
deriving Eq.~(20) we used the definition Eq.~(8) of the radion
field. The quantity $\rho$ is useful for our purpose because it
shows that the variations of Yukawa coupling constants (and other
radion-relating quantities we shall consider later) depend only on
the volume of the extra space, irrespective of how many extra
dimensions there are.

In the standard model, the Fermi constant $G_{\mathrm{F}}$ is not
a real fundamental constant \cite{Dixit1988}; rather, it could be
expressed as:
\begin{equation}
G_{\mathrm{F}} = \frac{g_{2}^{2}}{M_{\mathrm{W}}^{2}} =
\frac{g_{2}^{2}}{g_{2}^{2}\langle H\rangle^{2}} = \frac{1}{\langle
H\rangle^{2}},
\end{equation}
in which $g_{2}$ is the coupling constant of the SU(2) gauge group
and $M\mathrm{_{W}}$ is the mass of the weak gauge boson. In the
extension of the standard model to UED scenario, relation Eq.~(21)
still holds, but with the Higgs field and weak gauge boson
replaced by their corresponding zero Kaluza-Klein modes (see
Appendix A) and $g_{2}$ by the effective 4-dimensional SU(2) gauge
coupling constant. Therefore we conclude that the Fermi constant
is unaltered by the time evolution of the radion field.

The effective gauge couplings could be read from Eq.~(16) as:
\begin{equation}
g^{2} =
\tilde{g}^{2}\mathrm{exp}\left[-\kappa\sqrt{\frac{n}{n+2}}\sigma\right].
\end{equation}
Eq.~(22) means that
\begin{equation}
\frac{\alpha_{i,\mathrm{BBN}}}{\alpha_{i,\mathrm{NOW}}} =
\rho^{-1},
\end{equation}
where  $i$ = 1, 2, 3 corresponds to the U(1), SU(2) and SU(3)
gauge groups respectively, and the frequently used coupling
constants $\alpha_{i}$ are related to $g_{i}$ by $\alpha_{i} =
g_{i}^{2}/4\pi$. The fine structure constant
$\alpha_{\mathrm{EM}}$ is obtained as a combination of the
electroweak couplings, $\alpha^{-1}_{\mathrm{EM}} =
5\alpha_{1}^{-1}/3 + \alpha_{2}^{-1}$. Since $\alpha_{1}$ and
$\alpha_{2}$ have the same radion dependence described by
Eq.~(23), we have:
\begin{equation}
\frac{\alpha_{\mathrm{EM, BBN}}}{\alpha_{\mathrm{EM, NOW}}} =
\rho^{-1}.
\end{equation}

Finally let us consider the influence of the radion evolution on
the strong coupling. The quantity more relevant to our calculation
is the QCD scale parameter $\Lambda_{\mathrm{QCD}}$, which is
defined by the relation
$\alpha_{\mathrm{S}}(\Lambda_{\mathrm{QCD}}) = \infty$. Since
$\alpha_{\mathrm{S}} = \alpha_{3}$, the change of the strong
coupling constant is given by Eq.~(23). To obtain a relation
between it and the change in $\Lambda_{\mathrm{QCD}}$, we use the
one-loop renormalization group equation for QCD governing the
running of $\alpha_{\mathrm{S}}$:
\begin{equation}
\alpha_{\mathrm{S}}^{-1}(E) = \alpha_{\mathrm{S}}^{-1}(E_{0}) -
\frac{1}{2\pi}\left[-11+\frac{2}{3}n_{\mathrm{F}}\right]
\mathrm{log}\frac{E}{E_{0}},
\end{equation}
where $E$ and $E_{0}$ are any energy scales at which the values of
$\alpha_{\mathrm{S}}$ are measured and $n_{\mathrm{F}}$ the number
of quark flavors lighter than $E$. We shall choose the specified
energy scale $E_{0}$ to be the weak $Z$ boson mass,
$M\mathrm{_{Z}}$ = 91.2 GeV . The present measured value of
$\alpha_{\mathrm{S}}$ at this energy scale is
$\alpha_{\mathrm{S}}(M\mathrm{_{Z}}) = 0.119 \pm 0.002$. Then the
value of $E$ at which $\alpha_{\mathrm{S}}^{-1}(E) \rightarrow 0$,
\emph{i.e.} $E =\Lambda\mathrm{_{QCD}}$, could be solved to be
\begin{equation}
\Lambda_{\mathrm{QCD}} = M_{\mathrm{Z}}
\left[\frac{m_{b}m_{c}}{M_{\mathrm{Z}}^{2}}\right]^{\frac{2}{27}}
\mathrm{exp}\left[-\frac{2\pi}{9\alpha_{\mathrm{S}}(M_{\mathrm{Z}})}\right].
\end{equation}
In deriving Eq.~(26) we have made the assumption that
$\Lambda_{\mathrm{QCD}}$ lies between the strange and charm quark
masses $m_{s}$ and $m_{c}$ \cite{Dent2003, Dent2003b}. Note that
unlike some previous works, we choose not to calculate the
variations of the gauge couplings from the variation of a unified
gauge coupling here, because we are concerned with the
time-evolutions of the effective 4-dimenional gauge coupling
constants at energy scales far below the inverse size of the extra
dimensions, beyond which the field theory should be higher
dimensional (or effectively the higher order KK modes should be
taken into account).

Using Eq.~(23) and Eq.~(26), we get the relation between
$\Lambda\mathrm{_{QCD,BBN}}$ and $\Lambda\mathrm{_{QCD,NOW}}$:
\begin{equation}
\frac{\Lambda\mathrm{_{QCD,BBN}}}{\Lambda\mathrm{_{QCD,NOW}}} =
\rho^{-\frac{2}{27}}\mathrm{exp}\left[\frac{2\pi}
{9\alpha_{\mathrm{S,NOW}}(M_{\mathrm{Z}})}(1-\rho)\right].
\end{equation}

\section{VARIATIONS OF QUANTITIES RELEVANT FOR BBN CALCULATION}

In this section we will briefly review the standard BBN theory
(for more details see, \emph{e.g.} \cite{Kolb1990, SKM1993,
Sarkar1996, Serpico2004}) and discuss how its predictions are
influenced by the variations of the fundamental constants
considered in the previous section. Since some of these effects
have been discussed in the existing literatures, we will not
present them in details here.

At very early times ($T \gg$ 1 MeV or $t \ll$ 1 s), the energy
density of the universe is dominated by the photons, neutrinos and
relativistic electron-positron plasma, with a negligible
contribution from the baryons (mainly protons and neutrons). These
particles scatter frequently and are kept in thermal equilibrium.
In addition, the rates of the weak interactions
\begin{eqnarray}
\nu_{e} + n &\leftrightarrow& p + e^{-}\nonumber\\
e^{+} + n &\leftrightarrow& p + \bar{\nu_{e}}\nonumber\\
n &\leftrightarrow& p + e^{-} + \bar{\nu_{e}},
\end{eqnarray}
are far greater than the expansion rate of the universe, which is
given by the Friedmann equation
\begin{equation}
H^{2} = \frac{8\pi G}{3}\varrho_{\mathrm{total}},
\end{equation}
where $\varrho\mathrm{_{total}}$ is the total energy density of
the universe. As a result, there is also a chemical equilibrium
among these particles so that the ratio between the number
densities of neutrons and protons is
\begin{equation}
\frac{n_{n}}{n_{p}} \approx
\mathrm{exp}\left[-\frac{m_{np}}{T}\right],
\end{equation}
in which $m_{np} \equiv m_{n}-m_{p}$ is the mass difference
between neutron and proton. Eq.~(30) is valid because in the
standard BBN theory there is no lepton asymmetry and because of
the charge neutrality of the universe \cite{Kolb1990}.

As the temperature drops, the rates of weak interactions Eq.~(28)
decrease and become unable to keep neutrons and protons in
equilibrium. This begins to occur at a temperature of $\sim2\
\text{MeV}$ \cite{SKM1993}. After that, the actual $n/p$ ratio
will still be decreasing due to the free neutron decay and strong
$n-p$ reactions, until finally (essentially) all the neutrons have
been processed into nulcei and their number density becomes
constant.

At least down to the temperature $\sim0.8\ \text{MeV}$ the nuclear
reactions are capable of keeping the nuclei $\text{D}$,
${}^3\text{H}$, $^{3}\text{He}$, $^{4}\text{He}$ in both kinetic
and chemical equilibrium or the nuclear statistical equilibrium
(NSE) with corresponding abundances \cite{Kolb1990}
\begin{widetext}
\begin{equation}
Y_{A} =
g_{A}\left[\zeta(3)^{A-1}\pi^{\frac{1-A}{2}}2^{\frac{3A-5}{2}}\right]
A^{\frac{5}{2}}\left(\frac{T}{m_{N}}\right)^{\frac{3}{2}(A-1)}
\eta^{A-1}Y_{p}^{Z}Y_{n}^{A-Z}
\mathrm{exp}\left[\frac{B_{A}}{T}\right],
\end{equation}
\end{widetext}
where $g_{A}$ is the number of degrees of freedom of the nuclear
species $(A, Z)$, $m_{N}$ the nucleon mass, $B_{A}$ the binding
energy of $(A, Z)$, $\eta$ the baryon-to-photon ratio and $\zeta$
the Riemann zeta function. Because $\eta$ is of order $10^{-10}$,
Eq.~(31) says that the equilibrium abundances of the light nuclei
are very small at high temperatures.

With the temperature decreasing further, the nuclear abundances
provided by the reactions fall short of that required to maintain
the equilibrium and depart from NSE, firstly for the heavier
nuclei and later for lighter ones. The departure from NSE leads
directly to the result that the back reaction rates become much
smaller than the forward ones and are essentially switched off
\cite{Kneller2003}. By $ T \sim 0.2$ MeV, only the deuteron
abundance is still held in NSE and the abundances for all other
composite nuclei are many orders of magnitude lower than their NSE
values; some complex nuclei have been synthesized but the amounts
are yet negligible. Then the heavier nuclei ($^{4}\mathrm{He}$,
$^{3}\mathrm{He}$, $^3\text{H}$) abundances begin to follow the
deuteron NSE value. Finally, the deuteron bottleneck is passed at
$T \sim 0.1$ MeV and a significant amount of deuterium is
produced; the free neutrons are mostly rapidly assimilated into
$^{4}\mathrm{He}$, in which process some $^{3}\mathrm{He}$ and
$^3\text{H}$ are synthesized as the reaction ashes. These produced
nuclei interact and lead to the production of tiny abundances of
$^{7}\mathrm{Li}$, $^{6}\mathrm{Li}$ and $^{7}\mathrm{Be}$, the
last of which is finally turned into $^{7}\mathrm{Li}$ by the
electron capture process.

At the end of BBN nearly all free neutrons are incorporated in
$^{4}\mathrm{He}$, and so the final $^{4}\mathrm{He}$ abundance is
very sensitive to the neutron-to-proton ratio at freeze-out and
the duration between the freeze-out and the commence of BBN (for
some neutrons decay in this period). According to the review
above, then, the final $^{4}\mathrm{He}$ abundance will be varied
if at the time of BBN we have nonstandard values for the
neutron-proton mass difference $m_{np}$, the weak interaction
rates and (or) the cosmic expansion rate, the last of which will
be present in the case of a varying gravitational constant or a
varying energy density of some fluid component in the universe. In
contrast, the final yields of other complex elements depend
strongly on the various relevant nuclear reaction rates. Thus we
have to make clear how all these crucial physical quantities may
be modified with the varying fundamental constants as discussed in
the above section.

\subsection{Neutron-proton Mass Difference}

The neutron-proton mass difference is given phenomenologically as
\cite{Ichikawa2002}
\begin{equation}
m_{np} = m_{n} - m_{p} = m_{d} - m_{u}
-\alpha_{\mathrm{EM}}m_{\mathrm{EM}}
\end{equation}
with $m_{d}$, $m_{u}$ and $\alpha_{\mathrm{EM}}m_{\mathrm{EM}}$
being respectively the down quark mass, up quark mass and
electromagnetic self energy difference, in which $m_{\mathrm{EM}}$
is determined by strong interactions and proportional to the QCD
scale $\Lambda_{\mathrm{QCD}}$. The quark masses are not exactly
measured but could be calculated knowing the electromagnetic
contribution as $\sim 0.76$ MeV and the mass difference $m_{np}
\sim 1.293$ MeV \cite{PDG2000} to be $\sim 2.053$ MeV. In the
present model, since all the quantities appearing in Eq.~(32) may
change as the radion field evolves and the variations of quark
masses come completely from variations of the Yukawa couplings,
the neutron-proton mass difference at the time of BBN is evaluated
as
\begin{equation}
m_{np,\mathrm{BBN}} \doteq
2.053\frac{y_{\mathrm{BBN}}}{y_{\mathrm{NOW}}} - 0.76
\frac{\alpha_{\mathrm{EM,BBN}}}{\alpha_{\mathrm{EM,NOW}}}
\frac{\Lambda_{\mathrm{QCD,BBN}}}{\Lambda_{\mathrm{QCD,NOW}}},
\end{equation}
which could be expressed in terms of $\rho$ by using Eqs.~(20),
(24) and (27).

The formulae Eqs.~(32) and (33) are frequently used while
considering $m_{np}$'s fundamental-constant-dependences,
\emph{e.g.} \cite{Ichikawa2002, Ichikawa2004, Yoo2003,
Kneller2003, BIR, NL}.

\subsection{Weak Interaction Rates}

The weak rates of the neutron-proton inter-conversions can be
estimated using the Fermi theory. The $n\rightarrow p$ and
$p\rightarrow n$ rates are the summation of the rates of three
forward and backward reactions in Eq.~(28) respectively and could
be expressed as \cite{Kolb1990}:
\begin{widetext}
\begin{eqnarray}
\Gamma(n\rightarrow p) = &A& \int_{1}^{\infty} d\epsilon
\frac{\epsilon(\epsilon-q)^{2}(\epsilon^{2}-1)^{1/2}}{\left
[1+\mathrm{exp}(-\epsilon z_{e})\right ]\left
\{1+\mathrm{exp}\left [(\epsilon -q)z_{\nu}\right ]\right
\}}\nonumber\\ &+& A \int_{1}^{\infty} d\epsilon
\frac{\epsilon(\epsilon+q)^{2}(\epsilon^{2}-1)^{1/2}}{\left
[1+\mathrm{exp}(\epsilon z_{e})\right ]\left \{1+\mathrm{exp}\left
[-(\epsilon +q)z_{\nu}\right ]\right \}},
\end{eqnarray}
\begin{eqnarray}
\Gamma(p\rightarrow n) = &A& \int_{1}^{\infty} d\epsilon
\frac{\epsilon(\epsilon-q)^{2}(\epsilon^{2}-1)^{1/2}}{\left
[1+\mathrm{exp}(\epsilon z_{e})\right ]\left \{1+\mathrm{exp}\left
[(q-\epsilon)z_{\nu}\right ]\right \}}\nonumber\\ &+& A
\int_{1}^{\infty} d\epsilon
\frac{\epsilon(\epsilon+q)^{2}(\epsilon^{2}-1)^{1/2}}{\left
[1+\mathrm{exp}(-\epsilon z_{e})\right ]\left
\{1+\mathrm{exp}\left [(\epsilon +q)z_{\nu}\right ]\right \}},
\end{eqnarray}
\end{widetext}
where we have used dimensionless quantities $q=m_{np}/m_{e}$,
$\epsilon=E_{e}/m_{e}$, $z_{\nu}=m_{e}/T_{\nu}$ and
$z_{e}=m_{e}/T$ with $m_{e}$ being the electron mass, $T_{\nu}$
and $T$ respectively the temperatures of the neutrinos and the
electromagnetic plasma. $A$ is a normalization factor determined
by the requirement that at zero temperature $\Gamma(n \rightarrow
p + e^{-} + \bar{\nu_{e}})=\tau_{n}^{-1}$ where $\tau_{n}$ is the
neutron lifetime. This means that:
\begin{equation}
A = \tau_{n}^{-1} \lambda(q)^{-1} \propto
G_{\mathrm{F}}^{2}m_{e}^{5}
\end{equation}
where
\begin{equation}
\lambda(q) = \int_{1}^{q} d\epsilon\  \epsilon
(\epsilon-q)^{2}(\epsilon^{2}-1)^{1/2}.
\end{equation}

We have shown above that the Fermi constant $G_{\mathrm{F}}$ is
independent of the radion field, and so the variations of weak
rates at BBN from their current values originate from the changes
of $m_{e}$ and $q$, or equivalently from the changes of the Yukawa
coupling, the fine structure constant and the QCD scale
$\Lambda_{\mathrm{QCD}}$. It is then straightforward to calculate
the neutron lifetime and weak interaction rates at different
values of $\rho$ with the aid of Eqs.~(20), (24), (27), (33) and
(36). In our calculation we have done the explicit numerical
integrations of Eqs.~(34) and (35) in the BBN code to take into
account the effects of an evolving radion. For this purpose $A$ is
obtained by firstly computing $A_{\text{NOW}}$ using Eq.~(37) and
the measured value of $\tau_{n,\text{NOW}}$, and then calculating
$A_{\text{BBN}}$ with the aid of Eq.~(36). We have not evaluated
explicitly the various corrections to the weak rates
\cite{Serpico2004}, which are left for future work; instead we use
the Wagoner's approximation, \emph{i.e.}, reducing the
$n\leftrightarrow p$ rates by $2\%$. The difference between these
two treatments is estimated by Lopez and Turner \cite{Serpico2004}
to be $\sim0.5\%$ for ${}^4\text{He}$, which lies well within its
observational uncertainty (see below).

\subsection{Expansion Rate of the Universe}

Because we are working in the Einstein frame, the gravitational
constant will not change.  However, the change of the electron
(and also positron) mass does induce a variation of the energy
density stored in the electron-positron plasma because the energy
density and pressure both depend on the electron mass
\cite{Fowler1964}:
\begin{widetext}
\begin{eqnarray}
\varrho_{e^{+}} + \varrho_{e^{-}} &=& \frac{2}{\pi^{2}}m_{e}^{4}
\sum_{n=1}^{\infty}
(-)^{n+1}\mathrm{cosh}(n\phi_{e})\frac{1}{nz_{e}}
\left[\frac{3}{4}K_{3}(nz_{e})+\frac{1}{4}K_{1}(nz_{e})\right],\\
p_{e^{+}}+p_{e^{-}} &=& \frac{2}{\pi^{2}}m_{e}^{4} \sum
_{n=1}^{\infty}
\frac{(-)^{n+1}}{nz_{e}}\mathrm{cosh}(n\phi_{e})\frac{1}{nz_{e}}K_{2}(nz_{e}),
\end{eqnarray}
\end{widetext}
in which $\phi_{e}$ is the electron chemical potential and $K_{i}$
($i$ = 1, 2, 3) are the hyperbolic Bessel functions
\cite{Abramowitz1964}. In the actual calculations we truncate the
infinite summations in Eqs.~(38) and (39) to retain only the first
5 terms as in \cite{Kawano1992}.

Since this energy density is an important part in the total energy
budget of the universe, we expect that the expansion rate will
change also according to Eq.~(29), though this effect is not
significant. Note that the variation of $\Lambda_{\mathrm{QCD}}$
will cause the masses of all the nucleons and nuclei to be
different from their present-day (\emph{i.e.} the standard BBN)
values, acting as another source of nonstandard energy density.
However, because the contribution of baryons to the total energy
density is negligible, here we simply ignore this effect.

\subsection{Nuclear Reaction Rates}

The nuclear reactions relevant to BBN could be divided roughly
into the charged-particle-induced and the neutron-induced ones,
the former of which are characterized by the Coulomb barrier
penetration since the two charged particles must overcome their
Coulomb repulsion to approach each other and interact. In
\cite{BIR} the authors analyze the possible influences of a
varying fine structure constant on the cross sections of these
charged-particle reactions. By writing the cross section as
\begin{equation}
\sigma(E) = \frac{S(E)}{E}
\mathrm{exp}\left[-2\pi\alpha_{\mathrm{EM}}Z_{1}Z_{2}
\sqrt{\frac{\mu}{2E}}\right],
\end{equation}
in which $S(E)$ is the astrophysical factor, a slowly varying
function of energy, $Z_{i}$ ($i$ =1, 2) are the charges of the
reacting nuclei and $\mu$ is the reduced mass of the system, they
assume that all the dependence on $\alpha_{\mathrm{EM}}$ lies in
the exponential term in Eq.~(40). Then various thermonuclear
reaction rates are calculated using the fitting formulae of Smith,
Kawano \& Malaney \cite{SKM1993} (hereafter SKM) and expressed in
terms of the variation of $\alpha_{\mathrm{EM}}$. Later, Nollett
\& Lopez \cite{NL} improve the work of \cite{BIR} by including
several effects not considered there. In the present work we use
the original proposal of \cite{BIR} and include the improvements
by \cite{NL} in evaluating the reaction rates.

In addition to the influences from $\alpha_{\mathrm{EM}}$, the
variation of the reduced mass $\mu$ in the exponent of Eq.~(40)
could also change the cross section. This is because, with the
same energy, a different particle mass implies a different
particle velocity, and, since the penetration depends on the
velocity, it consequently implies a different possibility of
penetrating the Coulomb barrier. Meanwhile, these cross sections
may be changed with the modified nuclear binding energies and a
few reactions have resonance terms which are considered to be
dependent on the strong couplings. However, lacking explicit
expressions for the coupling dependence of these effects, we
choose to neglect them as in \cite{Ichikawa2002} and treat the
main contribution as coming from the exponent in Eq.~(40); an
estimation of the errors by neglecting these effects will be
presented in Sec.~VI (it is reassuring to observe the fact that,
since nuclear masses can be well approximated as proportional to
$\Lambda_{\mathrm{QCD}}$ and so $\mu\propto\Lambda_{\text{QCD}}$,
$\mu$ has a similar varying trend as that of
$\alpha_{\mathrm{EM}}$ and in fact several or many times larger in
magnitude. Thus the exponent in Eq.~(40) may be rather
significant). The thermonuclear rates taking into account the
$\Lambda_{\mathrm{QCD}}$ dependence are not difficult to obtain in
forms similar to those in \cite{BIR}, using the thermo-averaging
method of Fowler, Caughlan \& Zimmerman \cite{Fowler1967}.

There may also be dependences of the binding energies of the
complex nuclei on the strong coupling, which are not understood
very well but will also modify the inverse reaction rates.
However, we have checked that these effects are typically rather
small and negligible (at least for our allowed range of parameters
given below in Section VI) except for the reaction $p(n,
\text{D})\gamma$, which will be discussed below.

In contrast to the charged-particle-induced reactions, for the
neutron-induced reactions there are no Coulomb penetrations, and
their strong coupling dependences should be considered in the
context of the theoretical calculations of the cross sections.
Consider the reaction $p(n, \text{D})\gamma$, the first step of
BBN. The cross section of this reaction is poorly determined
experimentally \cite{Serpico2004}. Fortunately we have a rather
good theoretical understanding for it and whenever a comparison is
possible, the measurements show excellent agreements with
theoretical calculations. In this work, we adopt the calculation
with the low energy effective field theory without pions
\cite{Chen1999, Rupak2000}. The authors compute and express the
$p(n, \text{D})\gamma$ cross section as a function of the deuteron
binding energy $B_{d}$, the nucleon mass $M_{N}$, the scattering
length in the singlet channel and so on, concentrating on the
energy range relevant for BBN. The rate we obtain by
thermo-averaging this theoretical cross section \cite{Chen1999} is
virtually identical to the data-fitting results of \cite{SKM1993,
Cyburt2004}. To estimate the fundamental constant dependence of
the rate we, following \cite{Ichikawa2002}, assume the parameters
which appear in this theoretical formula and have dimensions
[Length]$^{n}$ to be proportional to $m_{\pi}^{-n}$. The deuteron
binding energy $B_{d}$ is important because it appears not only in
the cross section formula but also in the rate of the reverse
reaction $\gamma(\text{D}, n)p$ (and in fact it is the latter that
is dominant). Recent studies of the pion-mass-dependence of the
nuclear potential show a strong variation of $B_{d}$ with the pion
mass $m_{\pi}$ \cite{Epelbaum2003, Beane2003}; particularly, in
contrast to previous results, these authors show that $B_{d}$ is
in fact a \emph{decreasing} rather than increasing function of
$m_{\pi}$. As noticed in \cite{Yoo2003, Muller2004}, although the
calculations in \cite{Epelbaum2003, Beane2003} have large
uncertainties, we could safely approximate the deuteron binding
energy $B_{d}$ by
\begin{eqnarray}
B_{d,\mathrm{BBN}} &=& B_{d,\mathrm{NOW}}\left[(r+1)
-r\frac{m_{\pi,\mathrm{BBN}}}{m_{\pi,\mathrm{NOW}}}\right]
\nonumber\\
\ &\equiv& g(m_{\pi,\mathrm{BBN}})
\end{eqnarray}
within the narrow range we are interested in, \emph{i.e}. a small
variation of $m_{\pi}$. In Eq.~(41) the parameter $r$ varies from
6 to 10 according to the calculations of \cite{Epelbaum2003,
Beane2003}. For the present work we choose $r = 10$
\cite{Beane2003}. We have checked the case of $r = 6$ and found
that there is only a $\sim0.05\%$ increase in the best fitting
value of $\rho$ while the features discussed below are exactly the
same.

To relate Eq.~(41) to the fundamental constants discussed in the
previous section, we use the Gell-Mann-Oakes-Renner relation
\cite{GMOR} for pion mass:
\begin{equation}
f_{\pi}^{2}m_{\pi}^{2} = (m_{u} + m_{d})\langle\bar{q}q\rangle
\Rightarrow m_{\pi}\propto m_{q}^{1/2},
\end{equation}
in which the coupling of the pion to the axial current $f_{\pi}$
and the quark condensate $\langle\bar{q}q\rangle$ are proportional
to $\Lambda\mathrm{_{QCD}}$ and $\Lambda^{3}_{\mathrm{QCD}}$
respectively and $m_{q}$ denotes quark mass. Note that in
\cite{Beane2003} the authors, while quoting their results in terms
of the pion mass $m_{\pi}$, were varying the ratio of the quark
mass to $\Lambda_{\mathrm{QCD}}$, and so they indeed showed a
range of $B_{d}$ vs $m_{\pi}$ which is effectively a function
\cite{Kneller2003} (using Eq.~(42)) as
\begin{equation}
B_{d,\mathrm{BBN}} =
\Lambda_{\mathrm{QCD,BBN}}g'\left[\sqrt{\frac{m_{q,\mathrm{BBN}}}
{\Lambda_{\mathrm{QCD,BBN}}}}\right],
\end{equation}
where the function $g'$ is related with function $g$ in Eq.~(41)
through a change of variable. In Ref.~\cite{Kneller2003} Eq.~(43)
is used to obtain the relation between $B_{d}$ and $m_{np}$, both
of which depend on the quantity $\Lambda_{\mathrm{QCD}}$; here we
adopt the same relation to estimate the variation of $B_{d}$ in
the present model. There is also a weak dependence of $B_{d}$ on
the fine structure constant $\alpha_{\mathrm{EM}}$ because of the
small electromagnetic contribution (0.018 MeV) to $B_{d}$
\cite{Pudliner1997}, and we also have included this effect in our
calculation.

\begin{figure*}[]
\includegraphics[scale=1.19]{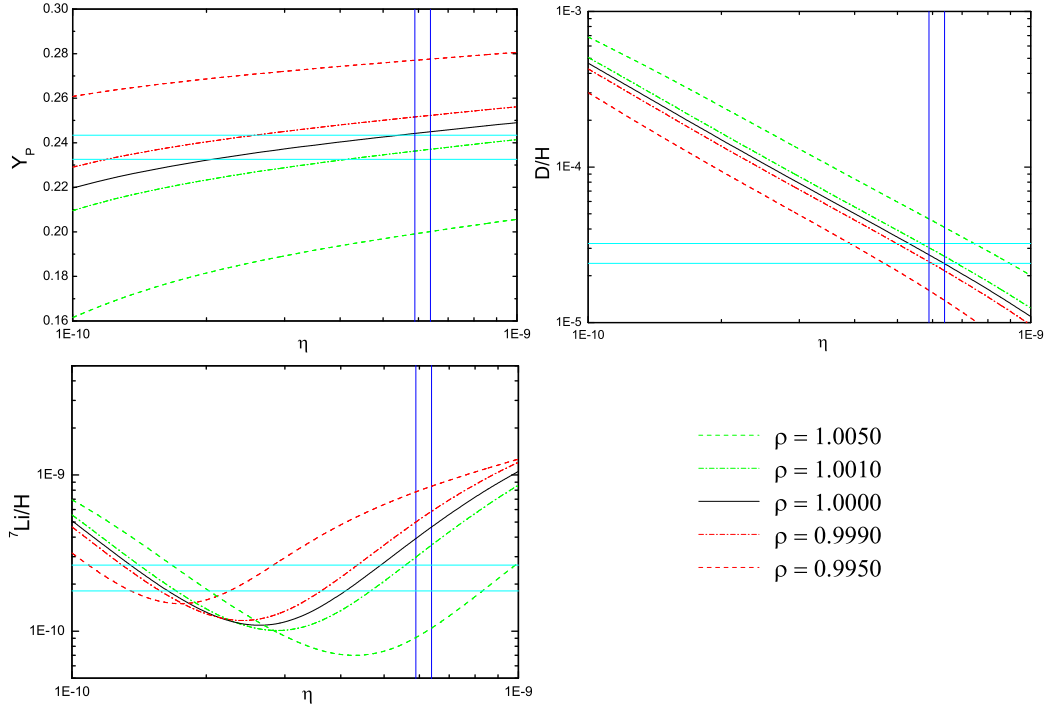}
\caption{\label{fig:epsart}(color online) The primordial
abundances of $^{4}\mathrm{He}$, D and $^{7}\mathrm{Li}$ as a
function of the baryon-to-photon ratio $\eta$, for various values
of $\rho$ as indicated by the legend in the lower right panel.
Also shown are the observational allowed ranges of these
abundances (the horizontal cyan lines) and the WMAP-indicated
range of $\eta$ (the vertical blue lines).}
\end{figure*}

\begin{figure*}[]
\includegraphics[scale=1.19]{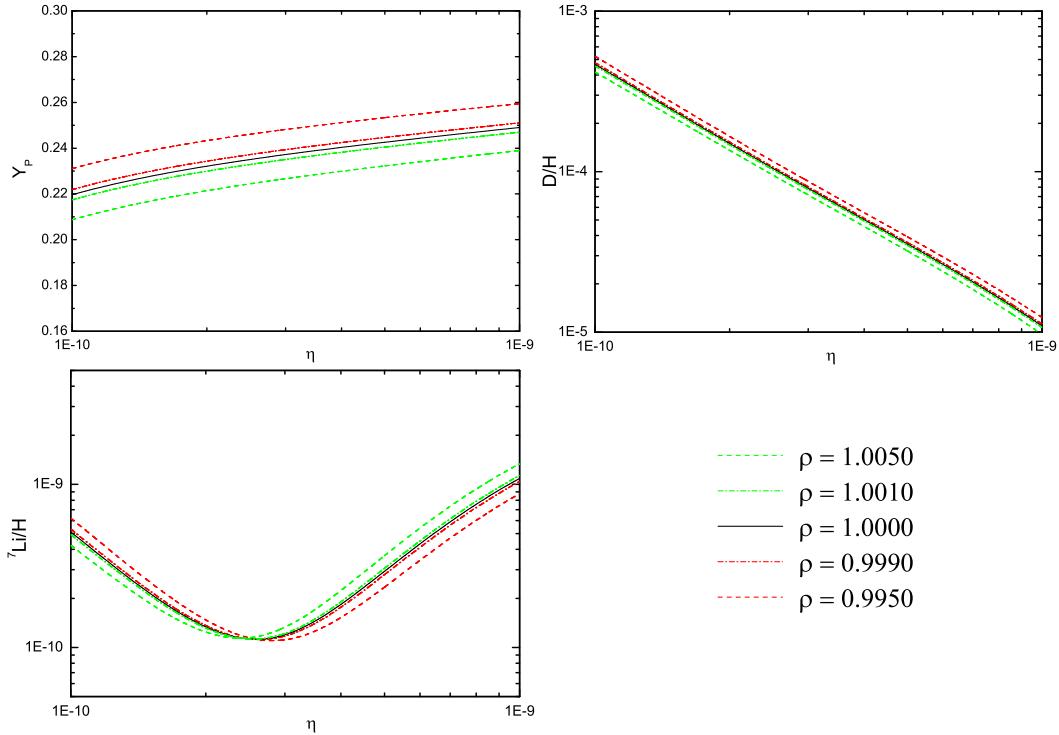}
\caption{\label{fig:epsart}(color online) Similar to FIG.1, but
here we neglect the effects coming from the reaction
$p(n,D)\gamma$ (mainly the deuteron binding energy $B_{d}$). See
the text for explanations.}
\end{figure*}

With the help of Eq.~(43) and the theoretical cross section for
$p(n, \text{D})\gamma$, it is then straightforward to obtain the
radion-dependence ($\rho$-dependence) of the thermo-averaged rate
for this reaction.

For the other two neutron-induced reactions $^{3}\mathrm{He}(n,
p){}^3\text{H}$ and $^{7}\mathrm{Be}(n, p)^{7}\mathrm{Li}$, there
is no theoretical calculation like that for $p(n,
\text{D})\gamma$, and so we simply use the data-fitting results
and neglect their coupling dependences. The reaction
$^{3}\mathrm{He}(n, p){}^3\text{H}$ influences the
$^{3}\mathrm{He}$ output which is not used to constrain the
parameters, and so the ignorance of its coupling dependences is
non-essential. But the $^{7}\mathrm{Li}$ abundance which we use to
reduce the parameter space does depend on the reaction
$^{7}\mathrm{Be}(n, p)^{7}\mathrm{Li}$ and this reaction exhibits
a broad resonance which may shift if the fundamental couplings
vary; in Sec.~VI we shall give a simple estimation on the error
due to neglecting its coupling dependences.

\section{NUMERICAL RESULTS}

We have incorporated all the above-discussed effects into the
standard BBN code by Kawano \cite{Kawano1992}. For the nuclear
reactions, we re-fit the updated rates given in the NACRE
compilation \cite{NACRE} (and a later work of Cyburt, Field \&
Olive \cite{CFO}) in the forms as in SKM to compute the impacts of
varying $\alpha_{\mathrm{EM}}$ and $\mu$. Then we compare this
result with that obtained by simply using the SKM-fitted rates and
find that they are essentially the same. Keeping this in mind, in
the following discussions we will only use the results derived
from the SKM rates.

In FIG.~1 we plot the abundances of the light nuclei
$^{4}\mathrm{He}$, D and $^{7}\mathrm{Li}$ including the effects
of an evolving radion field, which is characterized by the ratio
between the extra space volumes at the time of BBN and at present,
$\rho$.  It is apparent that if the size of the extra dimensions
is larger at the BBN era $(\rho > 1)$, then there will be an
increase in the deuterium output and decreases in the
$^{4}\mathrm{He}$ and $^{7}\mathrm{Li}$ (for $^{7}\mathrm{Li}$ we
only consider the larger-$\eta$-case in consistent with the WMAP
result \cite{Spergel2003}) yields compared with the standard BBN.
The behavior of the $^{4}\mathrm{He}$ output is mainly the
consequence of two effects originated respectively from the weak
interactions and the nuclear reaction $p(n, D)\gamma$. On one
hand, a $\rho$-value larger than 1 implies a smaller
$\alpha_{\mathrm{EM}}$ at the time of BBN, leading to a larger
$m_{np}$ (from Eq.~(33)) and smaller neutron density at
freeze-out, thus finally to a smaller $^{4}\mathrm{He}$ output
(note that as discussed above, the dependence on the reduced mass
$\mu$ has the same trend and strengthens this effect). Meanwhile,
the weak interaction rates will be larger than those in the
standard case, again decreasing the final $^{4}\mathrm{He}$
abundance. On the other hand, the larger $\rho$-value will cause a
smaller deuteron binding energy $B_{d}$ (from Eq.~(43)) than in
the standard BBN, and so the nucleosynthesis will commence at a
later time and become less efficient, producing less
$^{4}\mathrm{He}$ and $^{7}\mathrm{Li}$ while leaving more D
unprocessed (there is an extra decrease in the forward rate of
$p(n, D)\gamma$ which is also due mainly to $B_{d}$, but we have
checked that its influence is small compared with the effect of
the later commencing BBN). Both of these two effects work in the
same direction for $^{4}\mathrm{He}$ so that the final
$^{4}\mathrm{He}$ yield is very sensitive to $\rho$. As for
$\mathrm{D}$ and $^{7}\mathrm{Li}$, the effects of a larger $\rho$
other than that of $B_{d}$ come mainly from the Coulomb barrier:
the charged-particle-induced reaction rates will be less
suppressed since both $\alpha_{\mathrm{EM}}$ and $\mu$ are smaller
in this case, consequently more $\mathrm{D}$ is processed and more
$^{7}\mathrm{Li}$ is produced. To see this point more explicitly,
we plot in FIG.~2 the BBN yields if all the effects discussed in
the previous section \emph{except} that of the $p(n, D)\gamma$
reaction are included. FIG.2 agrees with the trend when only the
effects of changing $\alpha_{\mathrm{EM}}$ are considered, such as
that in \cite{BIR, Ichikawa2004}. However, the inclusion of
varying $B_{d}$ (as in FIG.~1) changes these abundances
dramatically, indicating the important role of the deuteron
binding energy in BBN \cite{Dmitriev2004, Kneller2003, Yoo2003}.

\begin{figure*}[]
\includegraphics[scale=1.19]{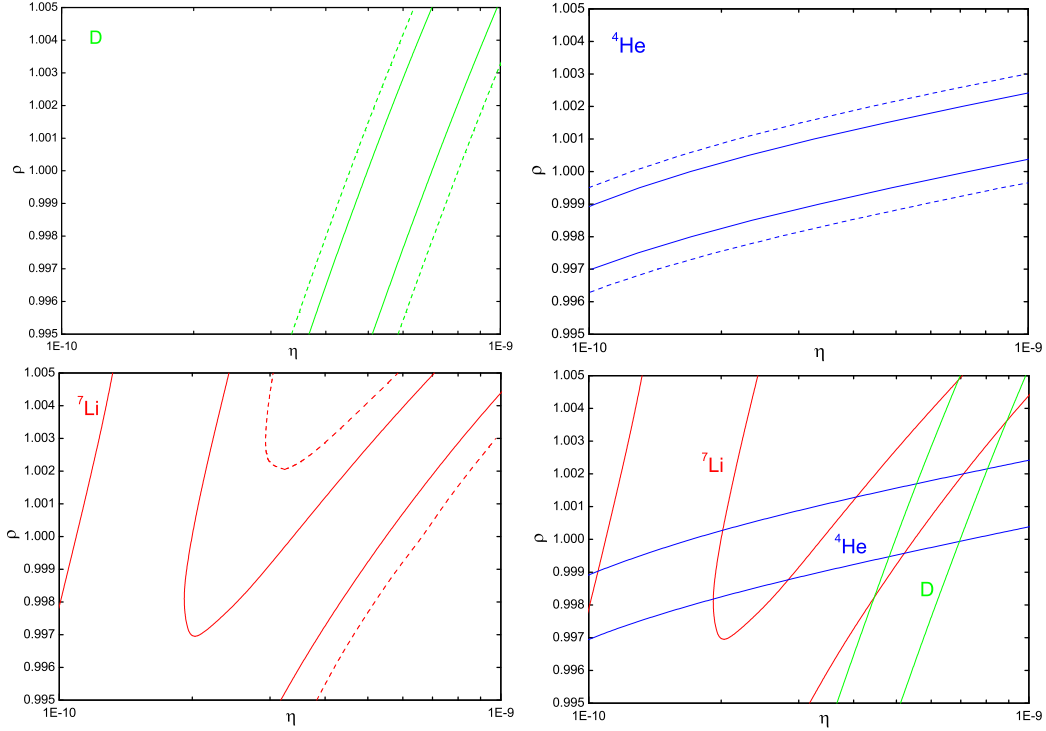}
\caption{\label{fig:wide}(color online) The 68\% (solid) and 95\%
(dashed) contours from the observational data on D (upper left),
$^{4}\mathrm{He}$ (upper right) and $^{7}\mathrm{Li}$ (lower
left). The lower right panel shows the 68\% contours plotted
together to make the comparison explicit.}
\end{figure*}

\begin{figure*}[]
\includegraphics[scale=1.25]{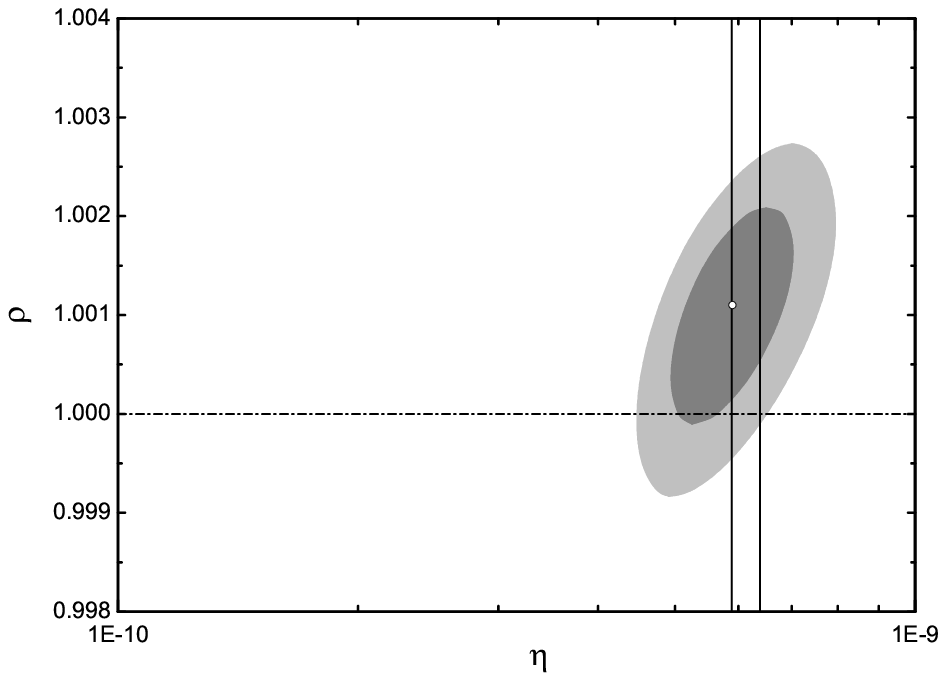}
\caption{\label{fig:wide}Joint constraint on the parameter space
from $\mathrm{D}$+$^{4}\mathrm{He}$+$^{7}\mathrm{Li}$; the gray
and light gray regions denote the 68\% and 95\% C. L. contours
respectively. The point with the maximum likelihood, which lies
around $(5.9\times 10^{-10}, 1.0011)$ on the $(\eta, \rho)$ plane,
is also shown (the white circle). The vertical lines represent the
$1\sigma$ range of $\eta$ determined by WMAP and the dot-dashed
horizontal line is the case of the standard BBN $(\rho=1)$.}
\end{figure*}

Our results as shown in FIG.~1 suggest that the discrepancy
between the standard BBN theory and WMAP observations tends to be
reduced if $\rho$ is greater than 1 \cite{Comment1}. Using the
WMAP-implied value of $\eta$ \cite{Spergel2003}, the standard BBN
can reproduce the observed D abundance, but not the
$^{4}\mathrm{He}$ and $^{7}\mathrm{Li}$ abundances; an $\eta$
smaller than the WMAP value is needed to obtain the observed
$^{4}\mathrm{He}$ and (or) $^{7}\mathrm{Li}$ abundances. While
this discrepancy may result from some systematic errors in the
$^{4}\mathrm{He}$ and $^{7}\mathrm{Li}$ observations, the
possibility of nonstandard BBN or new physics cannot be excluded.
In fact, there have been many attempts of solving this discrepancy
from a theoretical aspect, for example, by nonstandard expansion
rate \cite{Barger2003a}, varying fine structure constant
\cite{NL}, and lepton asymmetry \cite{Barger2003b} \emph{etc}.
These effects separately cannot solve the whole inconsistency and
a combination of them is needed, \emph{e.g.} \cite{Ichikawa2004}
(there are also works which use a single quantity to account for
the inconsistency, such as a varying deuteron binding energy
\cite{Dmitriev2004} and the Brans-Dicke cosmology with a varying
$\Lambda$ term \cite{Nakamura2005}). But from a theoretical
viewpoint, such combinations may not be necessary since, as
mentioned in the introduction, the variation of one fundamental
constant will often be accompanied by the variations of some other
constants; these changes together would play the role of several
effects combined. Because the co-variation of the fundamental
constants is rather model-dependent, a general study is not
practical. Our present work serves as a specific example and shows
how the variations of several fundamental constants are correlated
with one parameter $\rho$ and driven by the same physics - in this
case, the variation of the volume of extra-dimensional space.

In FIG.~1 the horizontal lines show the $1\sigma$ ranges of the
measured $^{4}\mathrm{He}$, D and $^{7}\mathrm{Li}$ abundances. We
use the observational results of Olive, Steigman and Walker
\cite{OSW2000} for $^{4}\mathrm{He}$ and the most recent work of
Kirkman \emph{et al.} \cite{Kirkman2003} for D:
\begin{equation}
Y_{\mathrm{P}}^{\mathrm{obs}} = 0.238 \pm 0.005,
\end{equation}
\begin{equation}
(\mathrm{D/H})^{\mathrm{obs}} = 2.78_{-0.38}^{+0.44} \times
10^{-5}.
\end{equation}
For Lithium we use the value given in the recent work of Bonifacio
\emph{et al.} \cite{Bonifacio2002},
\begin{equation}
(^{7}\mathrm{Li}/\mathrm{H})^{\mathrm{obs}} =
2.19_{-0.38}^{+0.46}\times 10^{-10},
\end{equation}
which is larger than previous results but still a factor of 2
smaller than the standard BBN + WMAP result. The vertical lines in
FIG.~1 represent the $1\sigma$ range of $\eta$ allowed by the CMB
analysis \cite{Spergel2003}:
\begin{equation}
\eta_{\mathrm{WMAP}} = (6.14 \pm 0.25) \times 10^{-10}.
\end{equation}

In order to see to what extent the quantity $\rho$ can differ from
1, we next carry out the likelihood analysis and find the
confidence contours in the two-dimensional parameter space $(\eta,
\rho)$. For this purpose, we adopt the semi-analytical method
introduced in \cite{Fiorentini1998} and then slightly generalized
in \cite{Cuoco2004, Serpico2004}. The likelihood function is
simply given as $L \propto \mathrm{exp}\left[-\chi^{2}/2\right]$
in which the chi-squared is calculated by:
\begin{equation}
\chi^{2} = \sum_{i,j}
\left[Y_{i}^{\mathrm{th}}-Y_{i}^{\mathrm{obs}}\right]W_{ij}
\left[Y_{j}^{\mathrm{th}}-Y_{j}^{\mathrm{obs}}\right],
\end{equation}
with $W_{ij}$ the inverse covariance matrix:
\begin{equation}
W_{ij} = \left[\sigma_{ij}^{2,\mathrm{th}}
+\sigma_{ij}^{2,\mathrm{obs}}\delta_{ij}\right]^{-1},
\end{equation}
and $Y_{i}^{\mathrm{th}}$ ($Y_{i}^{\mathrm{obs}}$) are the
theoretical (observational) abundances of the \emph{i}-th element.
In Eq.~(49) $\sigma_{ij}^{2,\mathrm{th}}$ is the error matrix
calculated by
\begin{equation}
\sigma_{ij}^{2,\mathrm{th}} = \frac{1}{4} \sum_{k}
\left[Y_{i}(\Gamma_{k}^{+})-Y_{i}(\Gamma_{k}^{-})\right]
\left[Y_{j}(\Gamma_{k}^{+})-Y_{j}(\Gamma_{k}^{-})\right],
\end{equation}
in which $\Gamma_{k}^{+(-)}$ is the \emph{k}-th nuclear reaction
rate plus (minus) its $1\sigma$ uncertainty; the summation is over
all the most relevant reactions. When $i=j$, Eq.~(50) simply gives
the theoretical uncertainty of the \emph{i}-th nuclear abundance.

In FIG.~3 we show the 68\% and 95\% C. L. contours on the
$\eta-\rho$ plane calculated using the modified Kawano code.
Although neither D nor $^{4}\mathrm{He}$ alone could constrain the
parameters considerably, we could see from the lower right panel
of this figure that the combination of them reduces the allowed
parameter space significantly, a similar conclusion as that in the
previous works \cite{Ichikawa2002,Yoo2003}. The WMAP-implied value
of $\eta$ is seen to lie in the allowed parameter space of anyone
of $\mathrm{D}$ \& $^{4}\mathrm{He}$, $\mathrm{D}$ \&
$^{7}\mathrm{Li}$ and $^{4}\mathrm{He}$ \& $^{7}\mathrm{Li}$. We
have also plotted the joint constraint on the parameters from
$\mathrm{D}$+$^{4}\mathrm{He}$+$^{7}\mathrm{Li}$ in FIG.~4 with
the WMAP-implied $1\sigma$ range for $\eta$, Eq.~(47), indicated
by the vertical lines. It can be seen that our best fitting value
(the white circle) lies just on the edge of the $1\sigma$ range of
$\eta_{\mathrm{WMAP}}$ while the whole $1\sigma$ WMAP range falls
into our 68\% C. L. contour. On the other hand, the standard BBN
$(\rho=1)$ cannot be consistent with the WMAP result and is
disfavored at the 68\% C. L. though it still lies inside our 95\%
C. L. contour. The allowed range of $\rho$ and $\eta$ by our
calculation is $0.9999 \lesssim \rho \lesssim 1.0021$,
$4.94\times10^{-10} \lesssim \eta \lesssim 7.05\times10^{-10}$ at
the 68\% C. L. and $0.9992 \lesssim \rho \lesssim 1.0027$,
$4.46\times10^{-10} \lesssim \eta \lesssim 7.99\times10^{-10}$ at
the 95\% C. L., of which the ranges of $\rho$ could be transformed
into the allowed variations of the fine structure constant in the
present model as:
\begin{widetext}
\begin{equation}
-2.1\times10^{-3}(-2.7\times10^{-3}) \lesssim
\frac{\alpha_{\mathrm{EM,BBN}}-\alpha_{\mathrm{EM,NOW}}}{\alpha_{\mathrm{EM,NOW}}}
\lesssim 0.1\times10^{-3}(0.8\times10^{-3}).
\end{equation}
\end{widetext}
This constraint is looser than that obtained in
\cite{Ichikawa2002} but more stringent than that in \cite{NL}.

\section{DISCUSSIONS AND CONCLUSIONS}

In summary, we have considered in this work the implications of a
slowly evolving radion field on the fundamental constants and the
Big Bang Nucleosynthesis predictions in the context of the
universal extra dimensions model. We have included in our modified
BBN code various effects of the varying constants, of which the
most important ones being the neutron-proton mass difference, the
deuteron binding energy, the free neutron lifetime and the nuclear
reaction rates. These effects may act in either the same or the
opposite directions and lead to distinct and complex behaviors of
the light nuclei yields. We have also calculated the chi-squared
as a function of $\rho$, the ratio between the extra space volumes
at the time of BBN and now, and $\eta$, the baryon-to-photon
ratio; this analysis then gives a bound on the variation of the
fine structure constant as $-2.1\times10^{-3} \lesssim
\Delta\alpha_{\mathrm{EM}}/\alpha_{\mathrm{EM}}\lesssim
0.1\times10^{-3}$ at 68\% C. L. and $-2.7\times10^{-3} \lesssim
\Delta\alpha_{\mathrm{EM}}/\alpha_{\mathrm{EM}}\lesssim
0.8\times10^{-3}$ at 95\% C. L. The allowed range for
$\Delta\alpha_{\mathrm{EM}}/\alpha_{\mathrm{EM}}$ in our result is
much larger than that obtained from quasar absorption systems
\cite{Murphy2001}, which is $
\Delta\alpha_{\mathrm{EM}}/\alpha_{\mathrm{EM}} = (-0.72 \pm
0.18)\times10^{-5}$. However, because BBN probes physics up to
redshifts of order $10^{10}$ while the quasar absorption systems
are at $z = 1 \sim 3$, there should be no apparent inconsistency.

It is suggested that a larger extra dimension size at the time of
BBN may help reduce the discrepancy between the BBN theory and
observation, and here we make several comments about this
conclusion. Firstly, the expression for the neutron-proton mass
difference $m_{np}$ (Eq.~(32)) is phenomenological but reasonable,
even though it may not be definitive \cite{BIR, NL}; we here use
it because it is the most-commonly-used and we find no competing
ones. Secondly, there is an uncertainty for the parameter $r$ in
the deuteron binding energy formula (Eq.~(41)) derived by the
effective field theory, but as we stated above, changing $r$ from
10 to 6 only induces a small variation of the numerical results
while having no influence on the qualitative features we have
discussed. Thirdly, our discussion on the variations of nuclear
reaction rates follows that in \cite{Ichikawa2002} and we have
neglected several effects due to the lack of theoretical
understandings for them. One example is the influence of a varying
$B_{d}$ on the reaction rates of
$\text{D}(p,\gamma){}^3\text{He}$,
$\text{D}(\text{D},n){}^3\text{He}$,
$\text{D}(\text{D},p){}^3\text{H}$,
$\text{D}({}^3\text{H},n){}^4\text{He}$,
$\text{D}({}^3\text{He},p){}^4\text{He}$, which are dominant or
subdominant in the destruction of $\text{D}$. Lacking an explicit
expression for this effect, we choose not to include it in the
modified code. Instead we can make an estimation about the error
introduced by neglecting it, by assuming the cross sections of
these reactions to scale with the deuteron radius and thus
$B_{d}$, \emph{i.e.}, $\sigma\propto1/(m_{N}B_{d})$
\cite{Kneller2003}. Adopting this parametrization in the code, we
find that the outputs of $\text{D}$, ${}^4\text{He}$ and
${}^7\text{Li}$ are changed by $-\lesssim9\%$, $\lesssim0.12\%$
and $\lesssim5\%$ respectively in our interested range of
$(\eta,\rho)$ (see FIG.~4), which lie well within the
corresponding $1\sigma$ observational uncertainties ($\sim16\%$
for $\text{D}$, $\sim2\%$ for ${}^4\text{He}$ and $\sim21\%$ for
${}^7\text{Li}$). Another example is the influence of resonances
in some key cross sections such as those of
$\text{D}({}^3\text{H},n){}^4\text{He}$,
$\text{D}({}^3\text{He},p){}^4\text{He}$ and
${}^7\text{Be}(n,p){}^7\text{Li}$. These resonant terms are
generally of the form \cite{Fowler1967}
\begin{equation}
\langle\sigma|v|\rangle_{\text{res}}=F(T)\text{exp}\left[-\frac{\bar{E}}{T}\right]
\end{equation}
in which $F(T)$ is a function of the temperature and $\bar{E}$ is
the resonance energy. Without a reliable calculation of how these
terms depend on the strong coupling, we roughly estimate the
corrections from them by assuming
$\bar{E}\propto\Lambda_{\text{QCD}}$. Then using the fitting
values of $\bar{E}$ in these cross sections \cite{SKM1993} in the
modified code we find that the $\text{D}$ and $^7\text{Li}$
abundances decrease by $\lesssim0.14\%$ and $\lesssim1.4\%$ in the
same range of $(\eta,\rho)$ as above, while the $^4\text{He}$
output stays essentially unchanged. These values again fall well
in the corresponding observational uncertainties. Although the
points discussed in this paragraph would not change our results
much, they do reflect the important role these effects play in
reducing possible errors, and in any case the precisions of the
BBN constraints will improve with future progress in particle and
nuclear physics relevant for these topics.

It is also interesting to look at whether the allowed range for
$\rho$ in our constraint would imply additional effects that are
too large to be dangerous for other observations, such as the
existence of bound di-proton/di-neutron and the stability of
${}^5\text{He}$. It is well known that the bound di-proton will
open a rapid channel for the hydrogen fusion \cite{Dyson1971} and
thus be catastrophic to the star lifetimes. The binding condition
for di-nucleon systems was considered by, \emph{e.g.~}Dent and
Fairbairn \cite{Dent2003}, and a rough criterion for these systems
to be stable was a decrease of $\Lambda_{\text{QCD}}$ to
$\sim1/10$ of its present day value (see \cite{Dent2003} for more
details), which is much larger than our constrained range for the
gauge coupling variations. The stability of $^5\text{He}$ is also
important in the consideration of varying fundamental constants
as, if stable, $^5\text{He}$ would fill the mass-5 gap in the BBN
nuclear reaction chain, drastically enhancing the production of
$^7\text{Li}$. This issue was discussed by, \emph{e.g.~}Flambaum
and Shuryak \cite{Flambaum2002}, and it was found there that the
limit from $^5\text{He}$ binding cannot compete with that from
$B_{d}$. For example, even the more stringent criterion to avoid a
stable $^5\text{He}$ quoted in Ref.~\cite{Flambaum2002}, namely
$\delta_{\pi}\equiv\delta(m_{q}/\Lambda_{\text{QCD}})
/(m_{q}/\Lambda_{\text{QCD}})\gtrsim-0.05$, is well satisfied by
our $2\sigma$ range of $\rho$.

As a final point, we comment on the choices of observational
$^4\text{He}$ and $^7\text{Li}$ abundances. The value given in
Eq.~(44) is taken from Ref.~\cite{OSW2000} and is certainly not
the only observational result of $Y_{\text{P}}$. Other data
analysis give different results, such as
$Y^{\text{obs}}_{\text{P}} =0.2421\pm0.0021$ in Ref.~\cite{IT2004}
and $Y^{\text{obs}}_{\text{P}} =0.249\pm0.009$ in
Ref.~\cite{Olive2004}. It is apparent from FIGs.~1 and 3 that, if
these larger abundances are used, the $^4\text{He}$ contour will
shift rightward. Meanwhile, as stated above, the $^7\text{Li}$
abundance as taken from Ref.~\cite{Bonifacio2002} is significantly
larger than previous estimates such as
$({}^7\text{Li}/\text{H})^{\text{obs}}=1.23\pm0.06^{+0.68}_{-0.32}$
(95\% C.~L.) given by Ryan \emph{et al.~}\cite{Ryan2000}. From
FIGs.~1 and 3 we see that, with the latter result adopted, the
right branch of the $^7\text{Li}$ contour (which we are interested
in) will shift leftward. Consequently the inconsistency between
BBN and WMAP as discussed above would be less reduced.

\begin{acknowledgments}
The work described in this paper was partially supported by a
grant from the Research Grants Council of the Hong Kong Special
Administrative Region, China (Project No. 400803). We thank
Dr.~F.~Iocco for providing a copy of the BBN code which we modify
and do the calculation with.  We also thank Dr.~J.~J.~Yoo,
Dr.~J.~P.~Kneller, Dr.~T.~Dent and Dr.~J.~D.~Barrow for helpful
information.
\end{acknowledgments}

\appendix

\section{A FIVE DIMENSIONAL TOY MODEL OF UED}

In this appendix we review briefly the main ingredients of the
minimal universal extra dimension model relevant for our
discussion. In particular we show that the UED model reduces to 4D
SM at low temperatures. For simplicity we will assume only one
extra dimension in this toy model, which is assumed to be flat and
compactified on an orbifold $\mathrm{S}_{1}/\mathrm{Z}^{2}$. The
fifth dimension is characterized by its coordinate $x_{4} = y \in
[0, 2\pi R]$, where $R$ is the radius of the fifth dimension (Note
that here the convention is different from that in Section II; see
\cite{Appelquist2001} for more details).

As a minimal generalization of SM, the 5-dimensional model
contains three generations of quarks and leptons, denoted by
$\tilde{Q_{i}}$, $\tilde{U_{i}}$, $\tilde{D_{i}}$, $\tilde{L_{i}}$
and $\tilde{E_{i}}$ respectively where $i = 1, 2, 3$ is the
generation index (For example, the zero modes of $\tilde{Q_{1}}$,
$\tilde{U_{1}}$, $\tilde{D_{1}}$, $\tilde{L_{1}}$ and
$\tilde{E_{1}}$ correspond respectively to the $(u,d)_{L}$,
$u_{R}$, $d_{R}$, $(\nu_{e},e^{-})_{L}$ and $e^{-}_{R}$ in the 4D
SM). Then the 5-dimensional Lagrangian in the matter sector could
be expressed as:
\begin{equation}
\tilde{L}_{5,\mathrm{total}} = \tilde{L}_{\mathrm{G}} +
\tilde{L}_{\mathrm{F}} + \tilde{L}_{\mathrm{H}} +
\tilde{L}_{\mathrm{Y}},
\end{equation}
in which the subscripts G, F, H and Y represent the gauge,
fermions, Higgs and Yukawa sectors of the model respectively and
are described as:
\begin{widetext}
\begin{eqnarray}
\tilde{L}_{\mathrm{G}} &=&
-\sum_{r=1}^{3}\frac{1}{2\tilde{g}_{r}^{2}}
\mathrm{Tr}\left[\tilde{F}_{r}^{AB}(X)\tilde{F}_{rAB}(X)\right],\\
\tilde{L}_{\mathrm{F}} &=& i\left[\bar{\tilde{Q}}(X),
\bar{\tilde{U}}(X), \bar{\tilde{D}}(X), \bar{\tilde{L}}(X),
\bar{\tilde{E}}(X)\right]\times\Gamma^{A}D_{A} \left[\tilde{Q}(X),
\tilde{U}(X), \tilde{D}(X), \tilde{L}(X),
\tilde{E}(X)\right]^{\mathrm{T}},\\
\tilde{L}_{\mathrm{H}} &=& \left[D_{A}\tilde{H}(X)\right]^{+}
D^{A} \tilde{H}(X) -\tilde{\mu}^{2}\tilde{H}^{+}(X)\tilde{H}(X) -
\tilde{\lambda}\left[\tilde{H}^{+}(X)\tilde{H}(X)\right]^{2},\\
\tilde{L}_{\mathrm{Y}} &=&
\tilde{Y}_{U}\bar{\tilde{Q}}(X)\tilde{U}(X)\cdot
i\sigma_{2}\tilde{H}^{\ast}(X) +
\tilde{Y}_{D}\bar{\tilde{Q}}(X)\tilde{D}(X)\tilde{H}(X)
\tilde{Y}_{E}\bar{\tilde{L}}(X)\tilde{E}(X)\tilde{H}(X)
+\mathrm{H.c.}.
\end{eqnarray}
\end{widetext}
In the above $\tilde{F}_{r}^{AB}$ denotes the 5-dimensional gauge
field strength associated with the
$\mathrm{SU}(3)_{\mathrm{C}}\times
\mathrm{SU}(2)_{\mathrm{W}}\times \mathrm{U}(1)_{\mathrm{Y}}$
group and $D_{A}$ the corresponding covariant derivative;
$\Gamma_{A}$'s are the 5-dimensional anti-commuting gamma matrices
defined as $\Gamma_{A} = \{\gamma^{\mu}, \gamma^{4}\}$ with
$\gamma^{\mu} \ (\mu = 0, 1, 2, 3)$ being the 4D Dirac matrices
and $\gamma^{4 }= - i\gamma^{0}\gamma^{1}\gamma^{2}\gamma^{3}$
(for the definition of gamma matrices in more than 5 dimensions
see \cite{Appelquist2001}). As in the standard model the fermion
masses are generated by the Higgs mechanism, the 5D Higgs doublet
$\tilde{H}(X)$ has the same form as in 4D, $\tilde{\mu}$ and
$\tilde{\lambda}$ being its potential parameters; the Yukawa
couplings $Y$'s are $3 \times 3$ matrices with mass dimension
$-1/2$. Notice that in Eq.~(A3) there is an implicit summation
over the 3 generations of quarks and leptons and that the capital
Latin letters A and B run from 0 to 4: $X^{A} = {x^{0}, x^{1},
x^{2}, x^{3}, y}$.

In the 4-dimensional effective theory the fields could be written
as an infinite summation of Kaluza Klein (KK) modes and the zero
KK modes are believed to play the role of the SM particles at low
energies. However, the KK expansion itself does not necessarily
lead to an exact equality of the zero KK modes to SM; to obtain
such an equality the KK decomposition should satisfy that, (1),
only left- (right-) handed component of each weak doublet
(singlet) is even under the orbifold projection and (2), the
redundant zero modes of the \emph{5th} components of gauge fields,
$\tilde{A}_{4}$ , should be eliminated. The KK expansions with
appropriate boundary conditions which satisfy the above
requirements are given as:
\begin{widetext}
\begin{eqnarray}
\tilde{H}(X)&=&\frac{1}{\sqrt{2\pi R}}\left[H^{(0)}(x) +
\sqrt{2}\sum_{n=1}^{\infty} H^{(n)}(x)
\mathrm{cos}\frac{ny}{R}\right],\\
\tilde{A_{\mu}}(X)&=&\frac{1}{\sqrt{2\pi R}}\left[A_{\mu}^{(0)}(x)
+ \sqrt{2}\sum_{n=1}^{\infty} A_{\mu}^{(n)}(x)
\mathrm{cos}\frac{ny}{R}\right],\\
\tilde{A_{4}}(X)&=&\frac{1}{\sqrt{\pi R}}\sum_{n=1}^{\infty}
A_{4}^{(n)}(x)\mathrm{sin}\frac{ny}{R},\\
\tilde{Q}(X)&=&\frac{1}{\sqrt{2\pi R}}\left\{Q_{L}^{(0)}(x) +
\sqrt{2}\sum_{n=1}^{\infty} \left[Q_{L}^{(n)}(x)
\mathrm{cos}\frac{ny}{R} + Q_{R}^{(n)}(x)
\mathrm{sin}\frac{ny}{R}\right]\right\},\\
\tilde{U}(X)&=&\frac{1}{\sqrt{2\pi R}}\left\{U_{R}^{(0)}(x) +
\sqrt{2}\sum_{n=1}^{\infty} \left[U_{R}^{(n)}(x)
\mathrm{cos}\frac{ny}{R} + U_{L}^{(n)}(x)
\mathrm{sin}\frac{ny}{R}\right]\right\},\\
\tilde{D}(X)&=&\frac{1}{\sqrt{2\pi R}}\left\{D_{R}^{(0)}(x) +
\sqrt{2}\sum_{n=1}^{\infty} \left[D_{R}^{(n)}(x)
\mathrm{cos}\frac{ny}{R} + D_{L}^{(n)}(x)
\mathrm{sin}\frac{ny}{R}\right]\right\},\\
\tilde{L}(X)&=&\frac{1}{\sqrt{2\pi R}}\left\{L_{L}^{(0)}(x) +
\sqrt{2}\sum_{n=1}^{\infty} \left[L_{L}^{(n)}(x)
\mathrm{cos}\frac{ny}{R} + L_{R}^{(n)}(x)
\mathrm{sin}\frac{ny}{R}\right]\right\},\\
\tilde{E}(X)&=&\frac{1}{\sqrt{2\pi R}}\left\{E_{R}^{(0)}(x) +
\sqrt{2}\sum_{n=1}^{\infty} \left[E_{R}^{(n)}(x)
\mathrm{cos}\frac{ny}{R} + E_{L}^{(n)}(x)
\mathrm{sin}\frac{ny}{R}\right]\right\}.
\end{eqnarray}
\end{widetext}
The factor $\sqrt{2}$ above is due to the different normalizations
of zero and higher order modes in the KK tower and will disappear
if we run the summation over both positive and negative values of
the KK numbers $n$. It's apparent from the above decompositions
that the zero modes are independent of the extra dimension
coordinate $y$, a fact which is expected because the standard
model should be purely 4-dimensional. The higher KK modes,
however, generally depend on $y$; and they will acquire additional
masses of order $n/R$, which will be accessible only with higher
enough energies because of the smallness of $R$ ($1/R$ should be
larger than several hundreds of GeV \cite{Appelquist2001}). Thus
up to the energy range relevant for BBN ($\mathcal{O}(1)$ MeV) the
world acts as 4-dimensional and the SM description shall be safe.

Lastly let us turn to the Higgs sector of this model. The fermion
masses are only generated after the spontaneous symmetry breaking
(SSB); after that, the Higgs field acquires a vacuum expectation
value which is determined by minimizing the Higgs potential
\begin{equation}
V(\tilde{H}) = \tilde{\mu}^{2}\tilde{H}^{+}\tilde{H} +
\tilde{\lambda}(\tilde{H}^{+}\tilde{H})^{2},
\end{equation}
in which $\tilde{\mu}^{2} < 0$ so that the potential has
nontrivial minima \cite{Oliver2003}. In the 4D effective theory
the mass of the \emph{n}-th Higgs KK mode is given by
$m_{H}^{(n),2} = \tilde{\mu}^{2} + n^{2}/R^{2}$. Therefore, if the
size of the extra dimensions is small enough such that
$|\tilde{\mu}|<1/R$, then $m_{H}^{(n),2}$ for $n \geq 1$ is
positive, and only the neutral zero KK mode obtains a nonzero
expectation value \cite{Oliver2003}. In this condition the zero
Higgs doublet KK mode could be expanded as:
\begin{equation}
H^{(0)} = \left(%
\begin{array}{c}
  \phi^{(0)+} \\
  \frac{1}{\sqrt{2}}(v+h^{(0)}+i\chi^{(0)}) \\
\end{array}%
\right),
\end{equation}
where $h^{(0)}$ is the physical Higgs zero mode $\phi^{(0)+}$ and
$\chi^{(0)}$ are the zero modes of Goldstone bosons; $v$ is the
(zero-mode) Higgs expectation value. At low energies as in our
case, the effects of the higher order Higgs KK modes are
negligible and will not be considered here.

% Create the reference section using BibTeX:
\newcommand{\noopsort}[1]{} \newcommand{\printfirst}[2]{#1}
  \newcommand{\singleletter}[1]{#1} \newcommand{\switchargs}[2]{#2#1}

\end{document}